\documentclass[twocolumn]{aastex701}
\usepackage{graphicx,lipsum, booktabs, footnote}
\usepackage{wrapfig}
\usepackage{float}
\usepackage{multirow}
\usepackage{graphicx, subfigure}

\usepackage[autostyle]{csquotes}

\begin{document}

\title{The SPOTLIGHT Pulsar Search Pipeline: A GPU-Accelerated FFT Approach}

\author[0009-0006-7995-5871]{Jyotirmoy Das}
\affiliation{National Centre for Radio Astrophysics (NCRA), Pune, 411007, Maharashtra, India}
\email{tataidas5392@gmail.com}

\author[0000-0002-2892-8025]{Jayanta Roy}
\affiliation{National Centre for Radio Astrophysics (NCRA), Pune, 411007, Maharashtra, India}
\email{jroy@ncra.tifr.res.in}

\author[0009-0001-4178-3879]{Nishant Pradeep Deo}
\affiliation{National Centre for Radio Astrophysics (NCRA), Pune, 411007, Maharashtra, India}
\affiliation{Indian Institute of Science Education and Research (IISER), Kolkata, Mohanpur-741246, West Bengal, India}
\email{nishantdeo8@gmail.com}

\author[0000-0003-2797-0595]{Karel Adamek}
\affiliation{Department of Physics, Silesian University in Opava, Opava, 74601, Czech Republic}
\email{karel.adamek@gmail.com}

\author[0000-0003-1756-3064]{Wes Armour}
\affiliation{Oxford e-Research Centre (OeRC), University of Oxford, Oxford-OX13PJ, United Kingdom}
\email{wes.armour@oerc.ox.ac.uk}

\author[0000-0002-8550-9070]{Kshitij Bane}
\affiliation{National Centre for Radio Astrophysics (NCRA), Pune, 411007, Maharashtra, India}
\email{kshitijbane@gmail.com}

\author[0009-0005-4130-892X]{Jayaram Chengalur}
\affiliation{Tata Institute of Fundamental Research (TIFR), Navy Nagar, Colaba, Mumbai-400005, Maharashtra, India}
\email{chengalur@ncra.tifr.res.in}

\author[0009-0003-8718-1560]{Kenil Ajudiya}
\affiliation{International Centre for Radio Astronomy Research (ICRAR), Curtin University, Bentley, WA 6102, Australia}
\email{kra635900@gmail.com}

\author[0009-0005-4130-892X]{Chahat Dudeja}
\affiliation{National Centre for Radio Astrophysics (NCRA), Pune, 411007, Maharashtra, India}
\email{dudejachahat11@gmail.com}

\author[0000-0003-3747-9847]{Sridhar Gajendran}
\affiliation{National Centre for Radio Astrophysics (NCRA), Pune, 411007, Maharashtra, India}
\email{sridhar.gajendran@gmail.com}

\author[0009-0002-6946-7541]{Zhaocheng Gong}
\affiliation{St Hilda's College, Oxford e-Research Centre (OeRC), University of Oxford, Oxford-OX13PJ, United Kingdom}
\email{zhaocheng.gong@st-hildas.ox.ac.uk}

\author[0009-0005-9293-0655]{Param Joshi}
\affiliation{National Centre for Radio Astrophysics (NCRA), Pune, 411007, Maharashtra, India}
\affiliation{Department of Physics and Astronomy, National Institute of Technology, Rourkela, 769008, India}
\email{paramjoshi1812@gmail.com}

\author[0000-0002-6631-1077]{Sanjay Kudale}
\affiliation{National Centre for Radio Astrophysics (NCRA), Pune, 411007, Maharashtra, India}
\email{kudale.sanjay@gmail.com}

\author[0009-0007-8409-4233]{Arpan Pal}
\affiliation{National Centre for Radio Astrophysics (NCRA), Pune, 411007, Maharashtra, India}
\email{apal@ncra.tifr.res.in}

\author[0000-0002-2441-4174]{Ujjwal Panda}
\affiliation{National Centre for Radio Astrophysics (NCRA), Pune, 411007, Maharashtra, India}
\email{ujjwalpanda97@gmail.com}

\author[0009-0006-6979-0655]{Adityan S}
\affiliation{Loyola College, Nungambakkam, Chennai-600034, Tamil Nadu, India}
\email{adityan.aero@gmail.com}

\author[0009-0002-2515-2425]{Raghav Wani}
\affiliation{National Centre for Radio Astrophysics (NCRA), Pune, 411007, Maharashtra, India}
\affiliation{Indian Institute of Science Education and Research (IISER) Pune, Pune-411008, Maharashtra, India}
\email{raghav.wani@gmail.com, wani.raghav@students.iiserpune.ac.in}

\author[0009-0003-7918-3208]{Junhao Zhang}
\affiliation{Department of Engineering Science, University of Oxford, Oxford-OX13PJ, United Kingdom}
\email{junhao.zhang@eng.ox.ac.uk}

\author[0009-0000-9152-3832]{Deepak Bhong}
\affiliation{National Centre for Radio Astrophysics (NCRA), Pune, 411007, Maharashtra, India}
\email{deepak@gmrt.ncra.tifr.res.in}

\author[0000-0000-0000-0000]{Shelton Gnanaraj}
\affiliation{National Centre for Radio Astrophysics (NCRA), Pune, 411007, Maharashtra, India}
\email{shelton@gmrt.ncra.tifr.res.in}

\author[0009-0007-4654-4539]{Santaji N. Katore}
\affiliation{National Centre for Radio Astrophysics (NCRA), Pune, 411007, Maharashtra, India}
\email{snk@gmrt.ncra.tifr.res.in}

\author[0009-0008-1233-6915]{Mekhala Muley}
\affiliation{National Centre for Radio Astrophysics (NCRA), Pune, 411007, Maharashtra, India}
\email{mekhala@gmrt.ncra.tifr.res.in}

\author[0000-0002-7551-5215]{Harshavardhan Reddy}
\affiliation{National Centre for Radio Astrophysics (NCRA), Pune, 411007, Maharashtra, India}
\email{reddysh@gmrt.ncra.tifr.res.in}

\begin{abstract}
We present the pulsar search component of SPOTLIGHT (Survey for sPoradic radiO bursTs via a commensaL multI-beam Gpu-powered Hpc at the gmrT), a GPU-accelerated commensal backend operating at the upgraded Giant Metrewave Radio Telescope (uGMRT). While SPOTLIGHT is primarily designed for real-time detection and localisation of fast radio bursts (FRBs), it simultaneously records a subset of beamformed data products for periodicity searches without requiring dedicated telescope time. To process the large data volumes generated by the survey, we have developed a scalable FFT-based pulsar search pipeline that combines radio-frequency interference mitigation, GPU-accelerated dedispersion and periodicity searches, multi-beam candidate sifting, efficient folding and machine-learning classification. Using population synthesis and archival uGMRT observations, we estimate that a fully operational SPOTLIGHT survey with 160 PC and one IA beam could discover $\sim$ 450 new pulsars, probing both high-sky coverage and faint pulsars over three and a half years of commensal observations. The pipeline has been validated on GMRT Cycle 48 and 49 observations (i.e. April 2025 to Mar 2026), successfully re-detecting numerous known pulsars with a wide range of Period, DM and flux densities, and is currently operational for SPOTLIGHT commensal data processing. We describe the SPOTLIGHT observing system, pulsar survey design, search parameter space, candidate-selection strategy, current status, and future developments. SPOTLIGHT demonstrates the scientific potential of commensal pulsar surveys and serves as a pathfinder for real-time, large-scale pulsar and transient searches in the SKA era.
\end{abstract}


\section{Introduction}
\label{sec:intro}

\makeatletter
\c@footnote=0
\makeatother

Following the discovery of the first pulsar in 1967 \citep{Hewish_1968}, extensive efforts were made to identify more pulsars and advance our understanding of these systems. Key discoveries--including the first binary pulsar \citep{Hulse_1975}, the first millisecond pulsar \citep{Backer_1982}, and the double pulsar system \citep{Burgay_2003}--significantly shaped the field of pulsar astronomy. These breakthroughs led to the development of major pulsar surveys. Early efforts such as the Parkes Multibeam Survey \citep{Manchester_2001, Lorimar_2006}, the Arecibo Pulsar survey \citep{Cordes_2006, Hessels_2007}, and the GBT Globular Clusters survey \citep{Ransom_2005} were followed by more recent ones, including the GBT Northern Celestial Cap survey \citep{Stovall_2014, McEwen_2024}, the GHRSS Pulsar Survey \citep{Bhattacharyya_2016, Singh_2023} and GCGPS \cite{Das_2025, Das_2026} with the uGMRT, the TRAPUM survey with MeerKAT \citep{Ridolfi_2022, Turner_2024}, and the ongoing, highly sensitive Five-hundred-meter Aperture Spherical Telescope (FAST) Galactic Plane Pulsar Snapshot (GPPS) survey \citep{Han_2021, Wang_2025}. Together, these surveys have yielded numerous discoveries, enabling extensive follow-up studies and driving major scientific advancements over the past decades. Pulsar discovery thus remains a key driver of progress in the field. To date, more than 4,300 pulsars have been identified, including about 700 millisecond pulsars\footnote{Data taken from the ATNF pulsar catalogue: \url{https://www.atnf.csiro.au/research/pulsar/psrcat/}}. This growing population has allowed for dense sampling of the parameter space $P$ - $\dot{P}$, revealing important statistical and evolutionary properties. However, continued discoveries remain crucial, as they not only improve this sampling but also can uncover previously unknown systems or properties, which allow us to study the system more thoroughly.

The total number of active radio pulsars in the Galaxy is estimated to be on the order of several hundred thousand \citep{Lyne_1985}, indicating that a large fraction remains undiscovered. With most of the sky already surveyed, current discoveries are increasingly sensitivity-limited—a trend clearly demonstrated by recent FAST results. This highlights the importance of improving survey sensitivity to detect fainter and more distant pulsars. Conducting a large-scale pulsar survey with such sensitivity requirements typically demands substantial telescope time and significant computational resources to process data efficiently and maximise discovery potential through wider sky coverage. However, allocating extensive dedicated telescope time is typically not feasible and can cause a compromise in other scientific studies. An effective alternative is to employ a commensal observing strategy, in which a dedicated backend \citep{Chennamangalam_2017, Surnis_2019} operates alongside ongoing observations without requiring additional telescope time. This approach enables rapid sky coverage while generating large volumes of data for pulsar and transient searches to do relevant science with it.

In this paper, we present such a pulsar search survey using the SPOTLIGHT\footnote{The SPOTLIGHT website: \url{https://spotlight.ncra.tifr.res.in/}} \citep{Roy_2024} backend. This commensal system has been developed for the uGMRT to operate in parallel with regular observations, enabling searches for pulsars, fast radio bursts (FRBs), and other radio transients. We first describe the backend and its operational aspects in Section~\ref{sec:spotlight}. The pulsar search project, including its motivation, design, and targeted science goals, is discussed in Section~\ref{sec:pulsar_search}. Finally, we summarise and conclude our results in Section \ref{sec:summary}.

\begin{figure*}
\centering
    \includegraphics[width=0.8\linewidth]{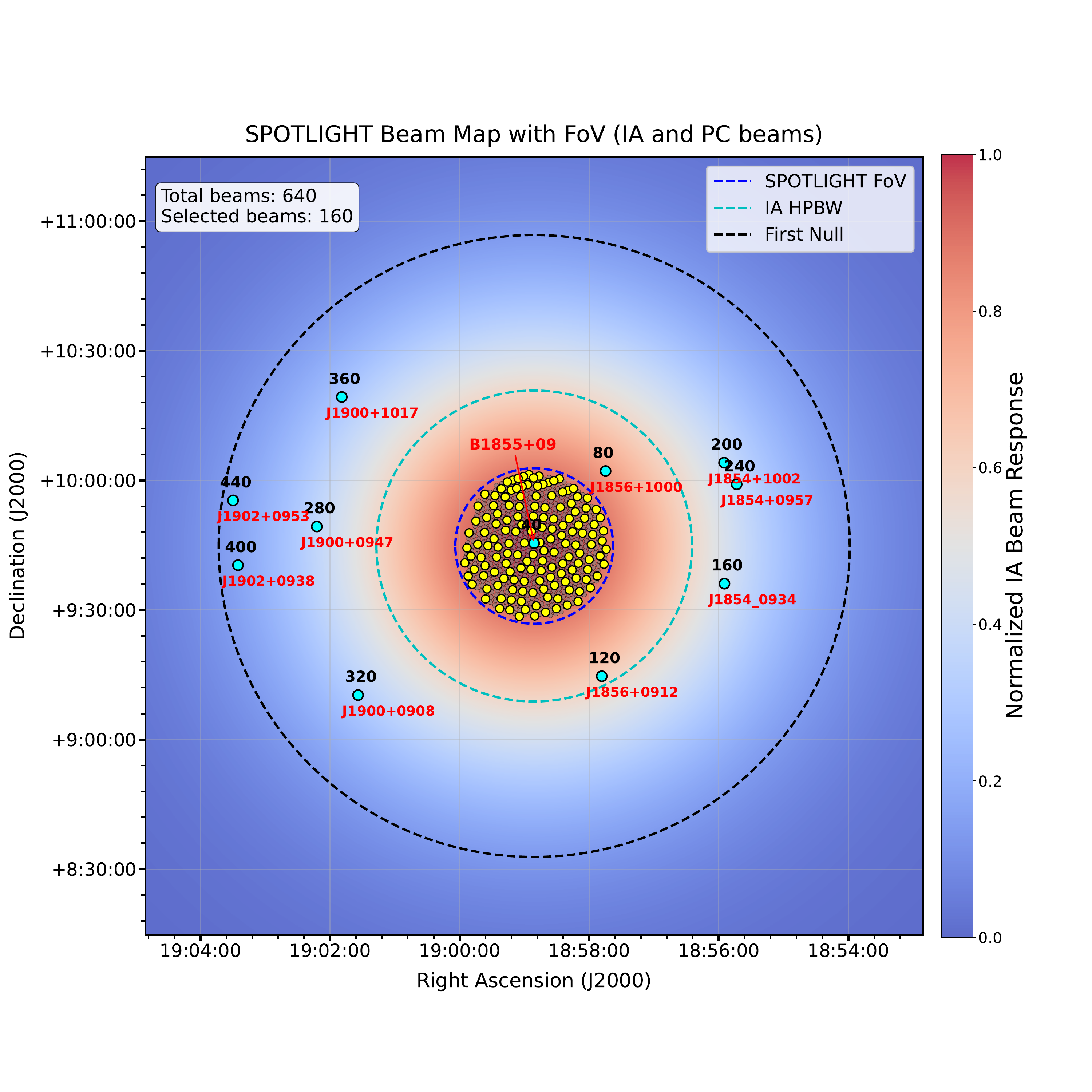}
    \caption{This figure shows the SPOTLIGHT beam formation for a typical commensal observation. A total of 640 PC beams (faint grey in the background, bright yellow and sky-coloured beams in the foreground), along with an IA beam (the colour plot), cover the whole Field of View (FoV) optimally. Out of the total 640 beams, some are targeted towards the known transient sources (pulsars/FRBs/RRATs) within the first null of the IA beam. Out of the total 640 beams, a total of 160 beams (bright yellow and sky colour beams) are dumped (including all the in-field sources targeted beams) for further pulsar search.}
    \label{fig:SPOTLIGHT_beam_tiling}
\end{figure*}

\section{SPOTLIGHT: The uGMRT Commensal Backend}
\label{sec:spotlight}
This section briefly describes SPOTLIGHT, the newly commissioned commensal search backend at the upgraded Giant Metrewave Radio Telescope (uGMRT). The system has been developed to enable high time-resolution and high-sensitivity sky surveys through a software-driven augmentation of the existing observing infrastructure. Operating in commensal mode, SPOTLIGHT performs wide-field pulsar and fast transient searches concurrently with primary observations, without compromising the scientific objectives of the main observing program.

\subsection{Overview}
\label{subsec:spotlight_overview}
SPOTLIGHT (Survey for sPoradic radiO bursTs via a commensaL multI-beam Gpu-powered Hpc at the gmrT) is a high-performance, GPU-accelerated backend developed to significantly enhance the scientific capabilities of the upgraded Giant Metrewave Radio Telescope (uGMRT). The system is designed to transform the uGMRT from a predominantly sensitive follow-up instrument into an efficient large-scale survey facility by enabling continuous, real-time processing of beamformed data products for pulsar and fast transient searches. For an overview of the SPOTLIGHT project, see \cite{Roy_2024}. Operating in a fully commensal mode, SPOTLIGHT maximises the scientific return of telescope time without interfering with the primary observational objectives.

The development and deployment of SPOTLIGHT were supported by the National Supercomputing Mission (NSM), Government of India, in collaboration with the Centre for Development of Advanced Computing (C-DAC), Pune. The data centre was commissioned at the uGMRT in October 2024.

The computational infrastructure comprises indigenously developed \textit{Param Rudra} servers hosting 90 NVIDIA A100 GPUs with 2880 CPUs, collectively referred to as \textit{Param Brahmand} facility, and is supported by a high-throughput storage system with an effective capacity of 2 petabytes (PB) for systematic archival of real-time commensal data products. The GPUs are distributed across 45 compute nodes, with each node hosting two GPUs along with 48 CPUs supporting both GPUs in each node. All these machines working together make the backend a state-of-the-art system, from data recording through analysis to the final results.

In the following sections, we provide a brief overview of the major components of the SPOTLIGHT backend, describing the signal-processing chain from the correlator and beamformer through to the data-recording system that enables subsequent pulsar-search processing.

\subsection{The correlator and beamformer}
\label{subsec:correlator_and_beamformer}
The SPOTLIGHT correlator and beamformer are designed to generate post-correlation beams \citep{Roy_2018} with a time resolution of 1.3 ms and 4096 frequency channels across the observing bandwidth. A detailed description of the SPOTLIGHT correlator and beamforming architecture will be presented in Reddy et al. (in preparation). In its commensal mode, during a scheduled uGMRT observation, the SPOTLIGHT system accesses the observation setup files, performs the initial phase calibration using the first calibrator scan, and generates the phasing solution required for subsequent beam formation on the target scans. At the beginning of each target scan, the beamformer module forms a total of (N) post-correlation (PC) beams (N=640 in the current implementation and up to N=2000 at full capacity), with adjacent beams spaced such that their responses overlap at approximately 40\% of the peak beam power. To maintain a balance between sensitivity and sky coverage, all PC beams are formed within the half-power beam width (HPBW) of the Incoherent Array (IA) beam. The number of arm antennas used in beam formation, which determines the beam size, is chosen accordingly, while all available central square antennas are included by default. This beam-steering cycle is repeated whenever a new calibrator scan is encountered, followed by the corresponding target scans, and continues until the observation is completed. We additionally form and record an Incoherent Array (IA) beam alongside the PC beams to further increase the surveyed sky area. This allows us to extend our sky coverage while simultaneously maintaining high-sensitivity coverage through the N post-correlation (PC) beams.

To further maximise the scientific return of the survey, a subset of the (N) post-correlation (PC) beams is dynamically assigned to known transient sources, including FRBs, pulsars, and RRATs, located within the first null of the telescope field of view. The positions of these sources are obtained from pre-loaded source catalogues maintained within the telescope operational framework, enabling simultaneous monitoring of known transients alongside blind survey observations.
A comprehensive overview of the real-time beam formation strategy is illustrated in Figure \ref{fig:SPOTLIGHT_beam_tiling} from a real observation. In this figure, the background colour map represents the recorded IA beam response, while the PC beams are tiled within the half-power beam width (HPBW) of the IA beam. Additionally, several PC beams are specifically targeted at known pulsars/FRBs/RRATs located within the first null of the IA beam. This simultaneous combination of wide-field coverage, high-sensitivity survey beams, and targeted observations of known transient sources highlights the versatility and scientific capability of the SPOTLIGHT beamforming system. 

Following beam formation, the SPOTLIGHT backend records a selected subset of 160 out of the 640 available PC beams, together with one IA beam, using 8-bit digitisation, a time resolution of 1.3 ms, and 4096 frequency channels across the full observing bandwidth. The complete set of 640 PC beams, along with the IA beam, is initially stored in a ring-buffer system and processed by the real-time FRB search pipeline. Subsequently, the selected 160-beam subset is written to disk for offline pulsar and transient searches.

\section{Pulsar Search with SPOTLIGHT: A GPU-Accelerated FFT Approach}
\label{sec:pulsar_search}
The primary objective of the SPOTLIGHT commensal backend is the real-time detection and localisation of FRBs, while simultaneously enabling periodic quasi-real-time searches for pulsars and other radio transients using the recorded beamformed data. While the detailed description of the FRB search pipeline is presented in \cite{Panda_2026}, in this section (i.e, paper), we focus on the FFT-based pulsar search pipeline, describing the scientific motivation for the survey, its discovery potential, the explored search parameter space, the end-to-end processing framework, and its current performance and status. We also summarise the initial results obtained from the ongoing observations, along with future developments in place.
 
\subsection{Motivation}
\label{subsec:motivation}
With more than 4000\footnote{Data taken from ATNF catalogue: \url{https://www.atnf.csiro.au/research/pulsar/psrcat/}} pulsars discovered to date, our understanding of pulsar physics and their applications in astrophysics has advanced significantly. However, population studies indicate that, in terms of discoverable pulsars, we have only begun to explore a small fraction of the total population. A large number of pulsars likely remain undetected, primarily due to limitations in survey sensitivity. Increasing sensitivity directly translates to the ability to detect fainter pulsars that are currently hidden below existing detection thresholds.

Discovering additional pulsars will allow for a more complete sampling of the $P$–$\dot{P}$ plane, improving our understanding of pulsar evolution and demographics. Moreover, there remains strong potential for uncovering rare and exotic systems, such as double neutron star (DNS) systems \citep{Tauris_2017}, highly eccentric binaries \citep{Champion_2008, Barr_2017}, and transitional pulsars \citep{Roy_2015}. Such systems are invaluable for follow-up studies and for testing fundamental physics. Despite the large number of known pulsars, the majority are classical pulsars, while exotic systems remain exceedingly rare. This strongly suggests that improved sensitivity could reveal a richer diversity of pulsar populations, including the possibility of discovering yet-unobserved systems such as MSP–MSP binaries or MSP–black hole binaries \citep{Barr_2024}.

Among the currently active pulsar surveys, the FAST Galactic Plane Pulsar Survey \citep{Han_2021, Han_2025} and the MPIfR-MeerKAT Galactic Plane Survey (MMGPS; \citealt{Padmanabh_2023, Colom_2023}) have contributed substantially to recent pulsar discoveries. FAST possesses a clear sensitivity advantage and operates over a frequency range that overlaps with that of the uGMRT. However, its sky coverage is limited to declinations above approximately $-17^\circ$, whereas the UGMRT can observe sources down to about $-54^\circ$. Conversely, MeerKAT provides excellent coverage of the southern sky, with a sensitivity comparable to that of the uGMRT (see Section 3.3 of \citep{Das_2025} for a detailed comparison). Together, these facilities provide complementary capabilities in sensitivity, frequency coverage, and sky accessibility, enabling efficient pulsar searches across a large fraction of the sky, particularly in the southern hemisphere.

These factors strongly motivate the use of the SPOTLIGHT backend on the uGMRT. Without requiring dedicated telescope time, SPOTLIGHT enables commensal pulsar searches. A subset of beams (see \ref{sec:spotlight} for a detailed discussion) from these observations can be stored and processed offline for pulsar searches. This approach provides an efficient pathway to conduct large-scale pulsar surveys, with the expectation of discovering a significant number of new pulsars from uGMRT commensal observations.

\begin{figure}
\centering
    \includegraphics[width=\columnwidth]{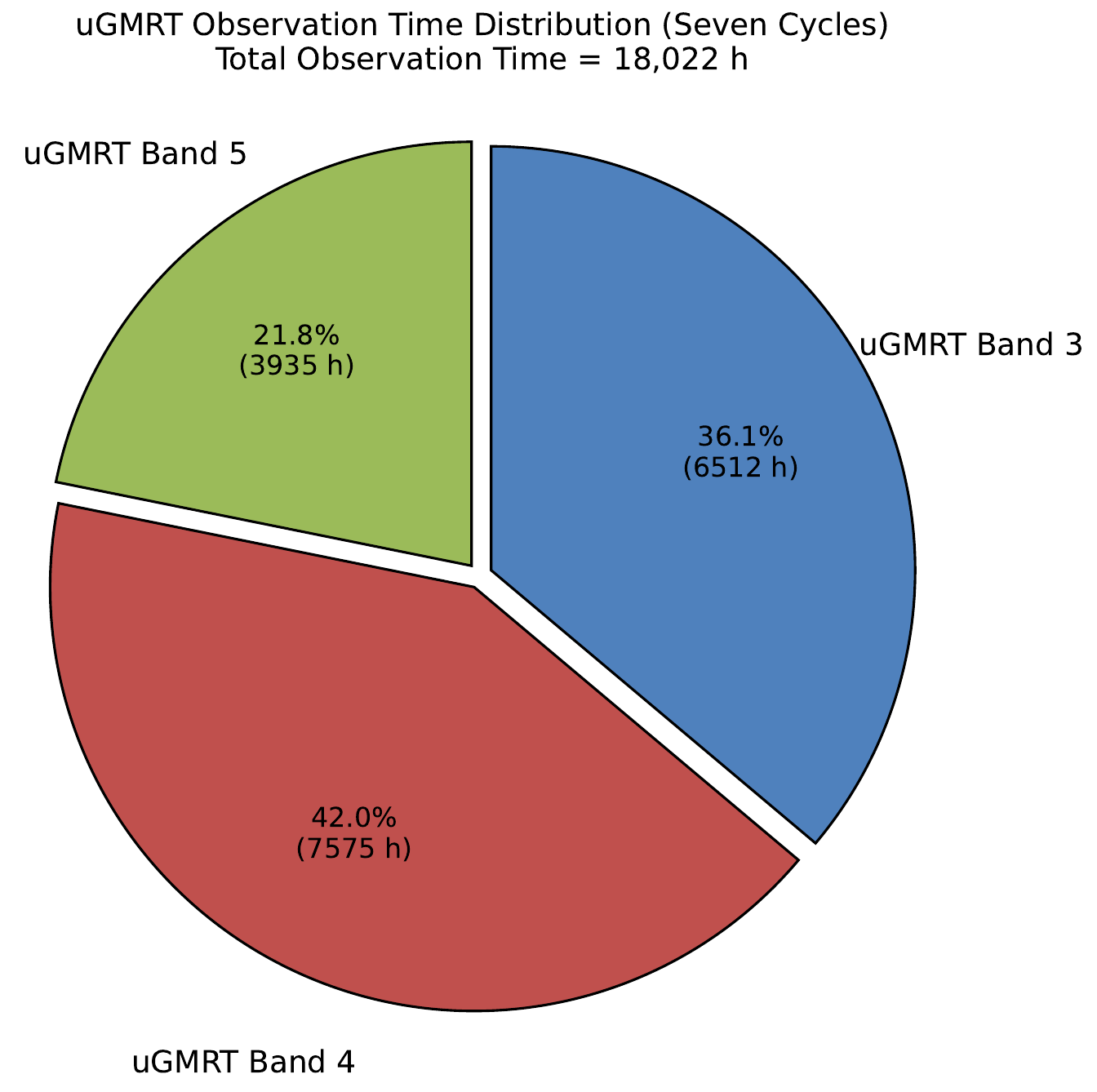}
    \caption{Distribution of the total uGMRT observation time across the three SPOTLIGHT commissioning bands (Band 3, Band 4, and Band 5), based on all target-source pointings obtained over seven uGMRT observing cycles. Calibrator scans are excluded. These observations constitute the dataset used for the pulsar discovery yield predictions presented in this work using {\tt PsrPopPy2}.}
    \label{fig:GMRT_observation_distribution}
\end{figure}

\begin{figure*}
\centering
\subfigure[Spin period vs Dispersion measure (DM) distribution]{
\includegraphics[width=0.48\textwidth]{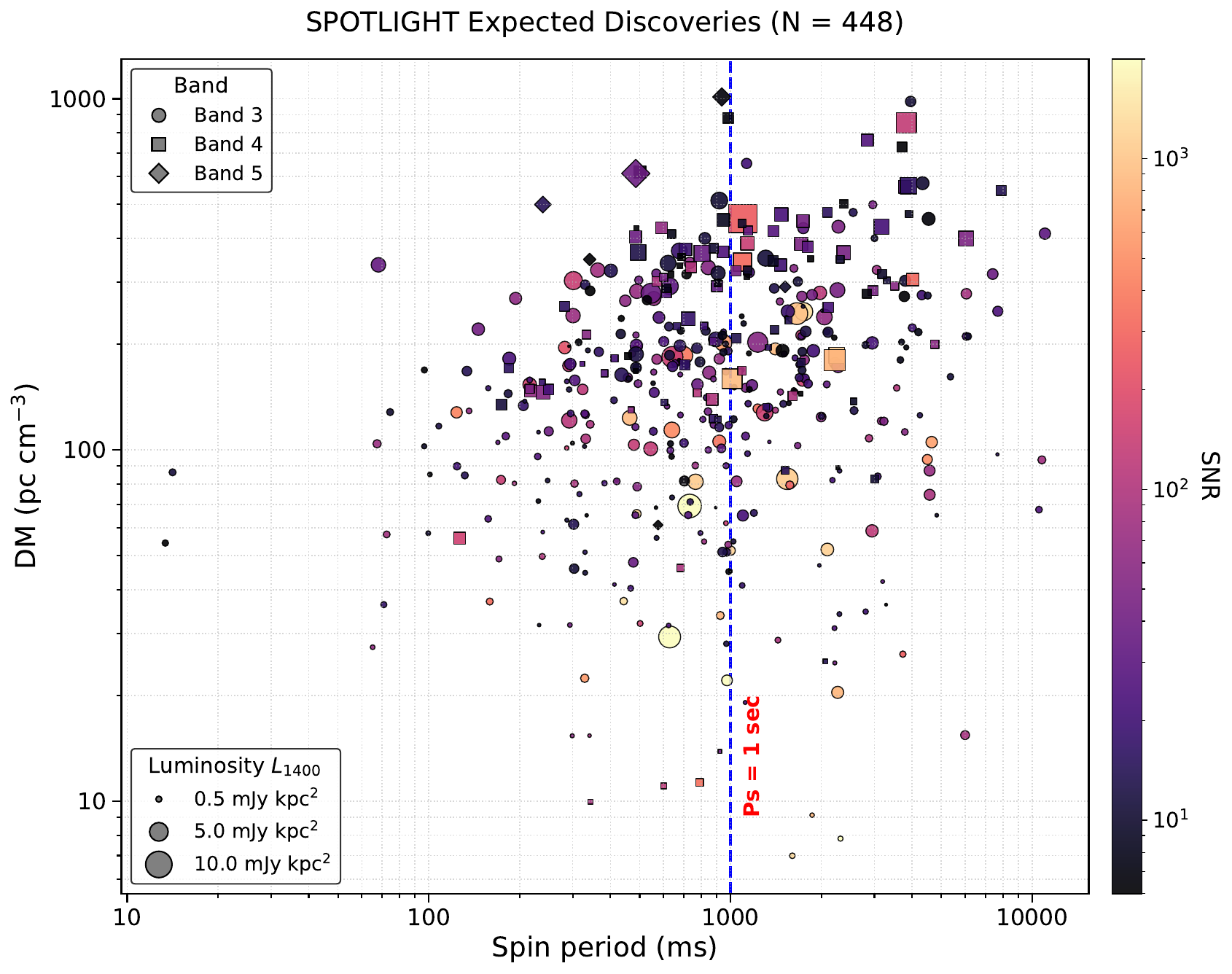}
}
\hfill
\subfigure[Spin period vs Flux density ($\rm S1400$) distribution]{
\includegraphics[width=0.48\textwidth]{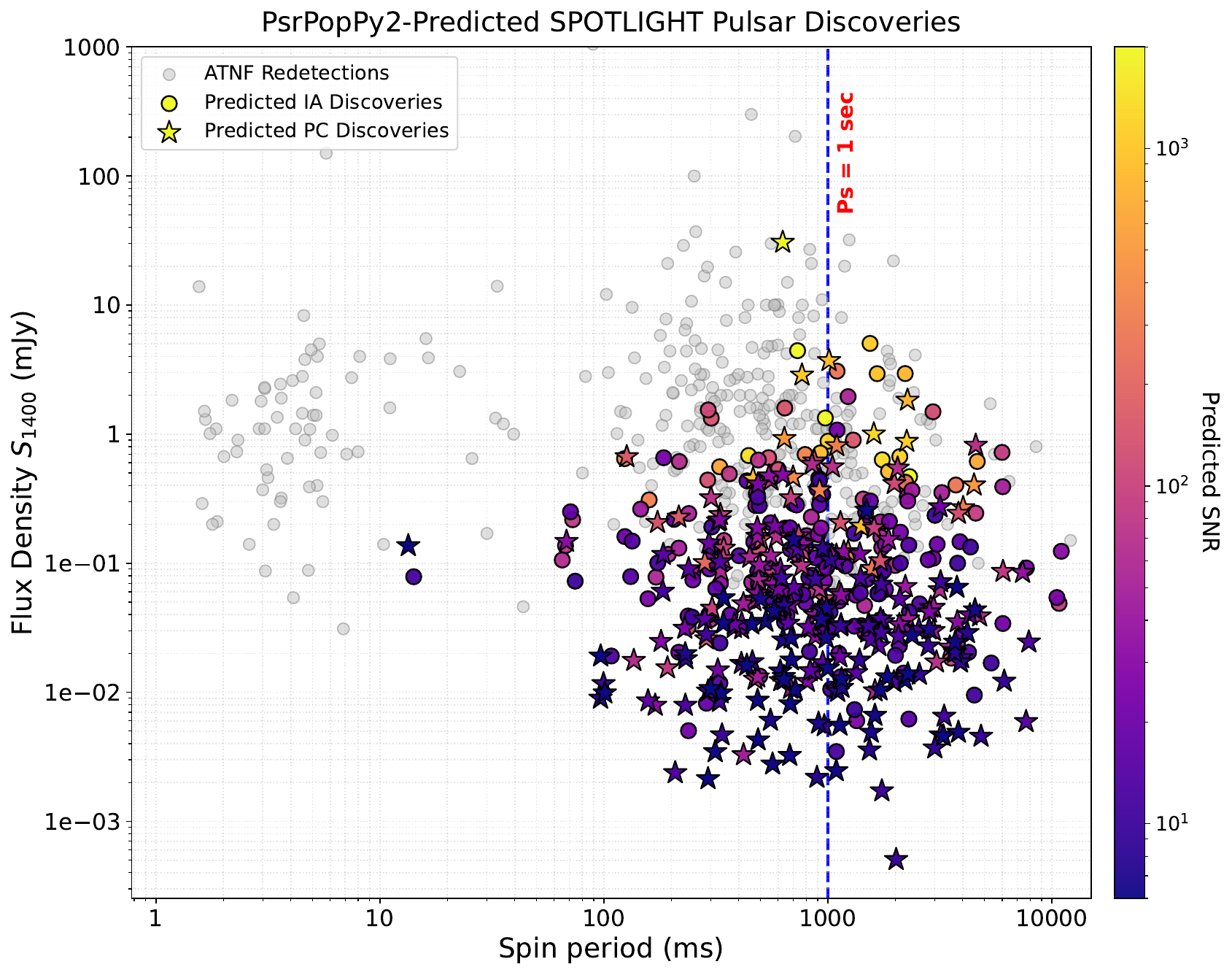}
}
\caption{Predicted properties of the 446 normal pulsars and 2 millisecond pulsars (MSPs) expected to be discovered by the SPOTLIGHT survey over three and a half years of operation using 160 PC beams and one IA beam, based on {\tt PsrPopPy2} population synthesis simulations over seven uGMRT cycles. (\textbf{a}) Dispersion measure (DM) versus spin period distribution of the predicted pulsar discoveries across the uGMRT observing bands. Marker colour represents the predicted signal-to-noise ratio (SNR), while marker size is proportional to the pseudo-luminosity ($L_{1400}$). (\textbf{b}) Flux density ($S_{1400}$) versus spin period distribution comparing the predicted SPOTLIGHT discoveries with ATNF pulsars redetected by the simulated survey. Grey circles denote ATNF redetections, coloured circles represent IA-predicted discoveries, and coloured star symbols represent PC-predicted discoveries. In both panels, the colour scale indicates the predicted SNR of the simulated discoveries.}

\label{fig:PsrPopPy2_outputs}
\end{figure*}

\subsection{SPOTLIGHT discovery prediction}
\label{subsec:discovery_prediction}
To predict the number of discoverable pulsars from SPOTLIGHT commensal observations, we use the pulsar population synthesis tool {\tt PsrPopPy2}\footnote{{\tt PsrPopPy2} GitHub repository: \url{https://github.com/devanshkv/PsrPopPy2}}. {\tt PsrPopPy2} is an updated version of the population synthesis framework {\tt PsrPopPy} \citep{Bates_2014}, incorporating improvements such as the inclusion of the Rotating Radio Transients (RRATs) population and the effects of interstellar scintillation, which were not available in the legacy version.

Using this synthesis framework, we first generate a snapshot pulsar population with the script {\tt populate.py}, adopting survey parameters from the Parkes Multibeam Pulsar Survey (PMSURV; \cite{Manchester_2001, Morris_2002}). PMSURV, conducted in the early 2000s using the Parkes radio telescope, discovered approximately 700 pulsars and significantly expanded the ATNF pulsar catalogue. We adopt PMSURV as the basis for our population synthesis because the pulsars expected to be detected in the SPOTLIGHT survey are anticipated to have similar properties. Using ATNF, we deduce how many pulsars PMSURV could detect with its survey parameters, and we use that detection in our snapshot {\tt ngen} to create our population basis. Two population basis were made for millisecond pulsars (MSPs with $\rm P_\mathrm{s} \leq 30\: ms$) and normal pulsars separately, with period and spectral index distributions drawn from the known ATNF population. Now, to estimate the sky coverage of the SPOTLIGHT commensal observations, we use three and a half years of archival uGMRT data (a total of seven uGMRT cycles). Since the SPOTLIGHT commensal recording system operates only in single-band mode (uGMRT Band-3, Band-4, and Band-5), we isolate all single-band observations from the archival dataset. Figure \ref{fig:GMRT_observation_distribution} shows the distribution of total observing time among the three bands. For each observation, we simulate the full set of 160 post-correlation (PC) beams and one additional IA beam to determine the effective sky coverage of the SPOTLIGHT commensal observations. To account for the random selection of 160 out of 640 PC beams and repeated observations of the same field, we incorporate the effective PC beams sky coverage into our simulations, with a limiting maximum coverage of 640 PC beams. Using this effective sky coverage, we employ the {\tt dosurvey.py} module of {\tt PsrPopPy2} to estimate the pulsar yield expected from the SPOTLIGHT survey, using the PMSURV population snapshot basis (both MSPs and normal pulsars) as the population basis. Now, each pointing assumes an effective sky coverage corresponding to the 160-PC-beam configuration (accounting for beam repetition) plus one IA beam, for the relevant uGMRT band, with sensitivity calculated using the integration time info from the archival scan length. For predicting the discovery, we used a minimum detection SNR threshold of $\sim \: 6\sigma$ for the much cleaner PC beams, and a conservative $\sim \: 10\sigma$ for the relatively higher radio-frequency interference (RFI) affected IA beam.

After combining detections across all pointings and removing redetections of already-known pulsars--identified via the ATNF pulsar catalogue and the SPOTLIGHT sensitivity limit--we predict that a fully operational SPOTLIGHT system with one IA and 160 PC beams could discover 448 new pulsars over three years of survey operations, combining IA and PC beam processing. Of these, 202 are independent IA detections, while the remaining 236 comprise either simultaneous IA and PC detections or independent PC detections. For consistency, all 236 are classified as PC detections, as the PC beam always yields a higher S/N than the IA beam in simultaneous detections. Figure~\ref{fig:PsrPopPy2_outputs} illustrates the properties (period, DM, SNR, and luminosity, etc.) of these predicted discoveries, along with the associated ATNF re-detections from the same pointings. Subfigure~(a) clearly shows the effectiveness of uGMRT Band~4 in detecting high-DM pulsars with intermediate sky coverage. The discovery prediction from Band~5 is comparatively lower, mainly because of the reduced sky coverage and the steep-spectrum nature of pulsar emission. Also, the wide distribution of spin period from $\sim$10~ms to $\sim$10~sec, with half of the discovery predictions above 1~sec, clearly stresses the necessity of different search methods to probe both slow and fast pulsars efficiently. In Subfigure~(b), a significant fraction of the detections arise from the IA beams owing to their wide field of view. In contrast, the PC beam, with its superior sensitivity, probes comparatively fainter pulsars, complementing the broad sky coverage provided by the IA beams and thereby enhancing the survey's discovery potential for very faint pulsars.

\begin{figure*}
\centering
    \includegraphics[width=0.8\linewidth]{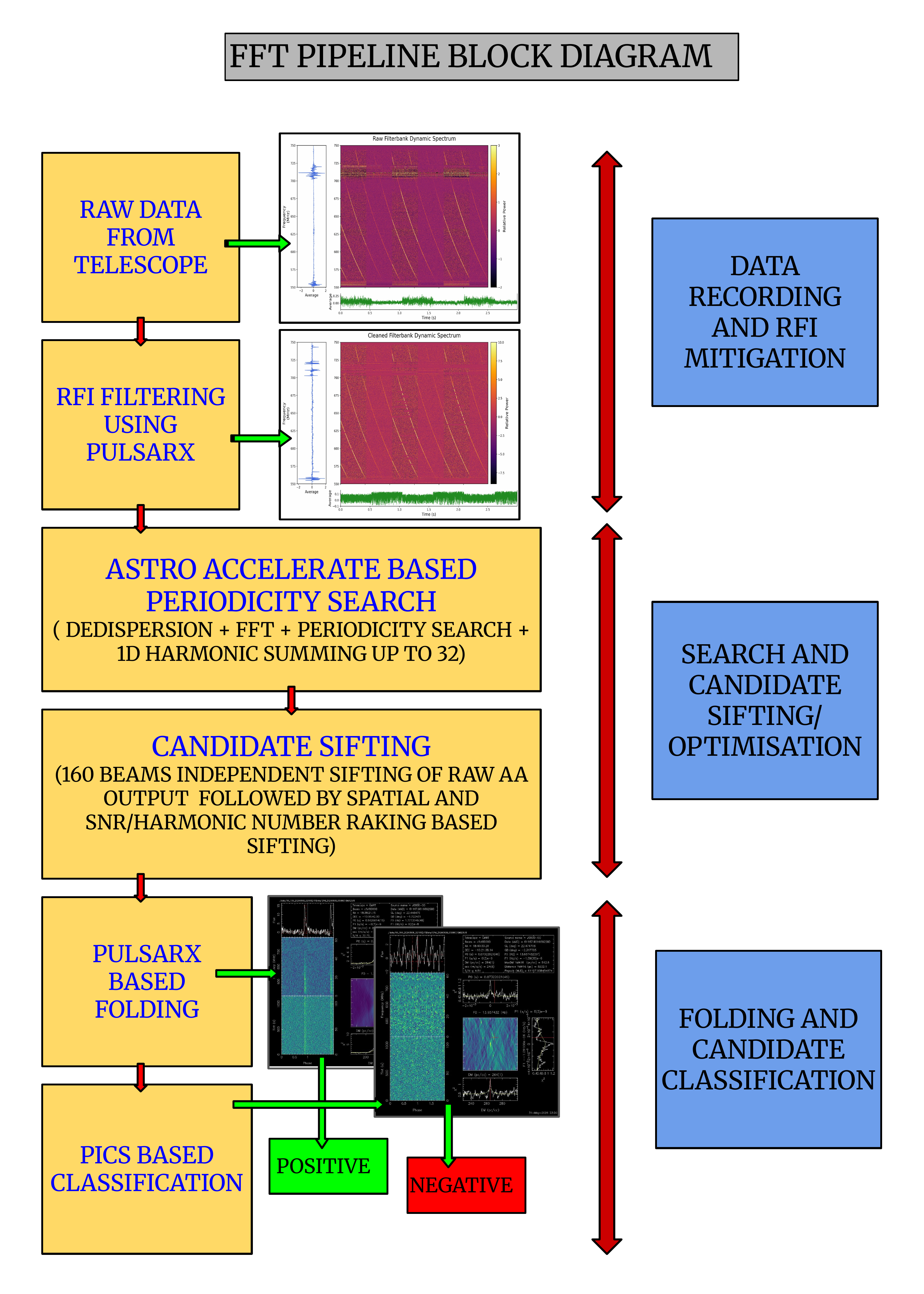}
    \caption{This figure outlines the main steps of the FFT pipeline ({\tt ver0}), starting from raw data recording and RFI cleaning, followed by an {\tt AstroAccelerate}-based acceleration search. Candidate sifting of multiple stages is performed, after which the final selected candidates are folded using {\tt PulsarX}. Finally, the folded pulse profiles are classified using the AI-based {\tt PICS} framework.}
    \label{fig:pipeline_block_diagram}
\end{figure*}

\subsection{The parameter space and search strategy}
\label{subsec:search_strategy}
The search parameter space is primarily determined by the time and frequency resolution of the incoming data. In the SPOTLIGHT commensal observing mode (see Section~\ref{sec:spotlight}), we record data with a time resolution of 1.3~ms and 4096 frequency channels across the observed bandwidth, which may be either 200 or 400~MHz depending on the observing configuration. These resolutions are guided by the primary science objective of real-time FRB searches, with their inherent property of short time-scale bursts from great distances (i.e. higher dispersion measure) needing both time and frequency relations to be as good as possible.

With a sampling time of 1.3~ms, the periodicity search becomes most sensitive to mildly recycled millisecond pulsars (MSPs; $P_\mathrm{S} \gtrsim 10$~ms; \cite{Berezina_2017}) and normal pulsars with spin periods from tens of milliseconds to seconds. Sensitivity to shorter periods is limited by finite sampling and dispersion smearing, while at longer periods, FFT searches become less effective because of dominant red noise, as seen in \cite{Cameron_2017}. Consequently, our FFT-based search is optimised and implemented for periods between $\sim$10~ms and 1~s. As presented in Section \ref{subsec:discovery_prediction} and Figure \ref{fig:PsrPopPy2_outputs}, a significant number of discoverable pulsars lie beyond $\rm P_\mathrm{s} = 1\, \mathrm{s}$. Now to detect these slower pulsars with improved sensitivity, we also developed a Fast Folding Algorithm (FFA) based pipeline (Bane et al. in preparation) covering periods from 1~s to several tens of seconds, where FFA methods outperform FFT searches in the long-period regime \citep{Singh_2023}. Together, these complementary pipelines provide sensitivity across the full range of periodicities accessible to SPOTLIGHT observations.

In addition to constraining the periodicity range, the input sampling time also places an effective upper limit on the dispersion measure (DM) that can be searched while maintaining optimal sensitivity. For a given observing band, intra-channel dispersion smearing and interstellar scattering limit us to how high in DM space we can search without significant signal degradation. This effect is most severe at lower observing frequencies, where dispersive delays scale more strongly with frequency. Within the SPOTLIGHT commensal system, the lowest observing band is the uGMRT band-3 (300–500 MHz), and therefore, this band imposes the most restrictive DM constraints. To determine the maximum DM for each band while maintaining manageable computational loads, we use the {\tt PRESTO}\footnote{{\tt PRESTO} GitHub repository: \url{https://github.com/scottransom/presto.git}} tool {\tt DDplan.py} and extend the DM search range until the intra-channel smearing matches the sampling time of 1.3 ms, with a maximum limit of 1000 $\rm pc\:cm^{-3}$. This yields maximum DM limits (instrumental limits) of approximately 250 pc cm$^{-3}$ for band-3, 800 pc cm$^{-3}$ for band-4 (550–950 MHz), and 1000 pc cm$^{-3}$ for band-5 (1060–1460 MHz). These limits are set by the time resolution, channelisation, and computational constraints of the search. However, since the actual interstellar electron content along a given line of sight is often significantly lower, searching the full instrumental DM range is frequently unnecessary. Therefore, to reduce computational cost even more without compromising discovery potential, we use the NE2025 Galactic electron density model (\citealt{Ocker_2025}) to estimate the maximum Galactic DM, $\mathrm{DM}^{*}$, for each target direction, assuming a fiducial distance of 25 kpc. The practical DM search range is then defined as the smaller of either $3\times\mathrm{DM}^{*}$ or the instrumental DM limit for the observing band. This approach maintains sensitivity to sources beyond the model prediction while avoiding unnecessarily large DM searches, thereby optimising both the DM plan and the overall processing time.

\begin{figure*}
\centering
\subfigure[Stage 0: All candidates from unfiltered raw AA output (1000X reduced)]{
\includegraphics[width=0.48\textwidth]{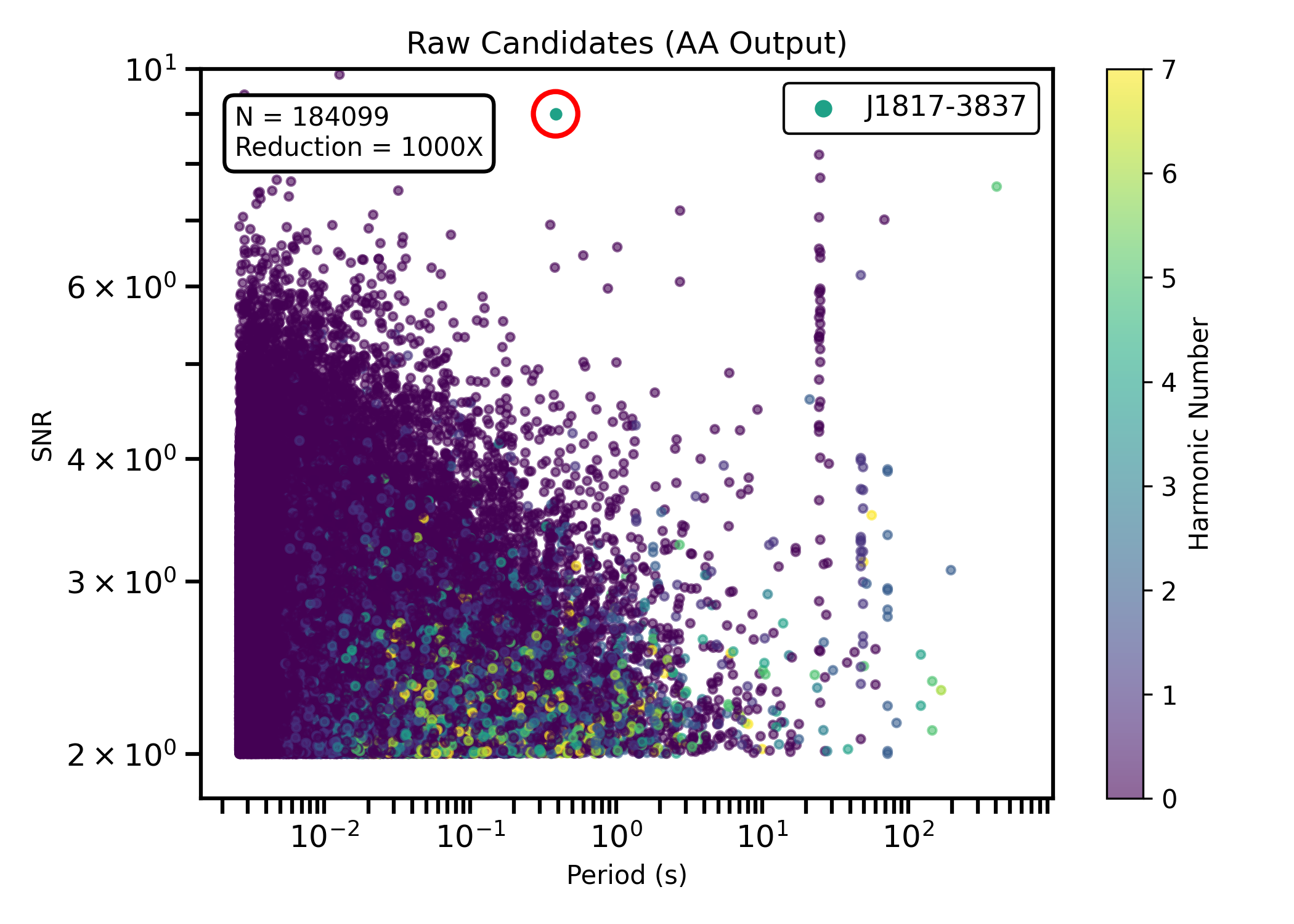}
}
\hfill
\subfigure[Stage 1: Candidates after independent sifting of each beams using DM clustering and SNR threshold (5X reduced)]{
\includegraphics[width=0.48\textwidth]{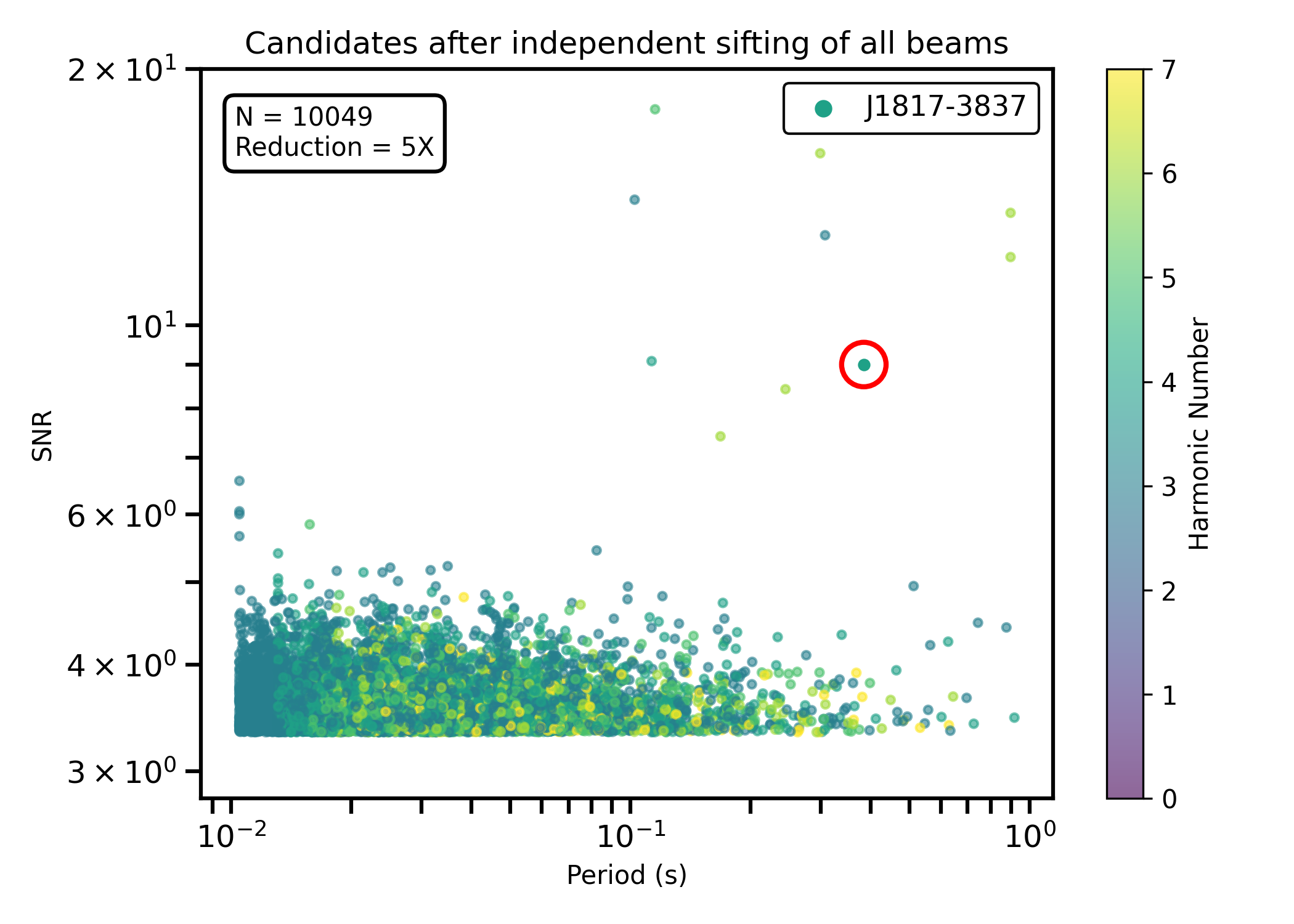}
}
\vspace{0.3cm}
\subfigure[Stage 2: Candidates after sifting based on spatial clustering and harmonic + SNR ranking]{
\includegraphics[width=0.48\textwidth]{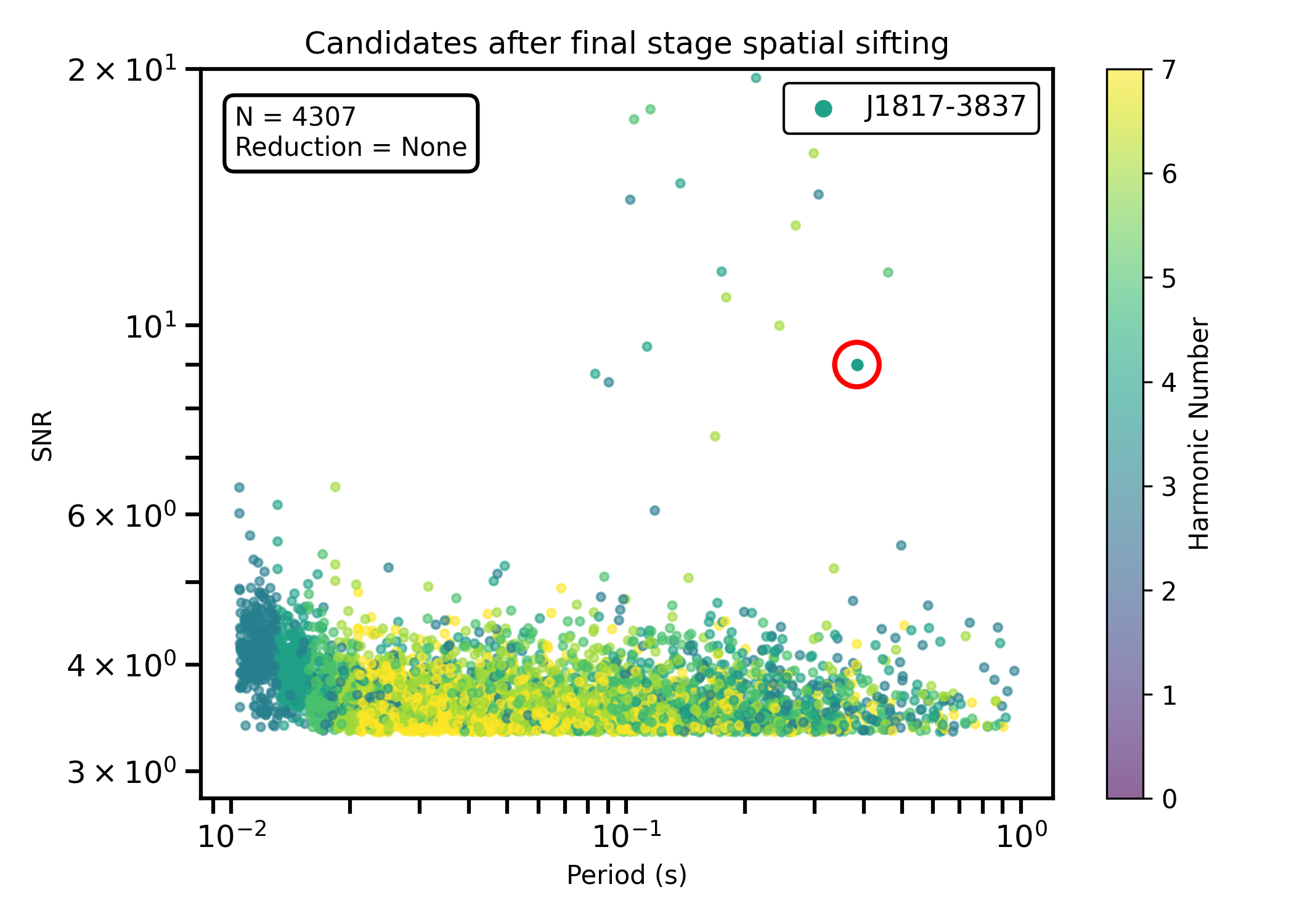}
}
\hfill
\subfigure[PICS classifier output for all folded candidates after final spatial sifting]{
\includegraphics[width=0.48\textwidth]{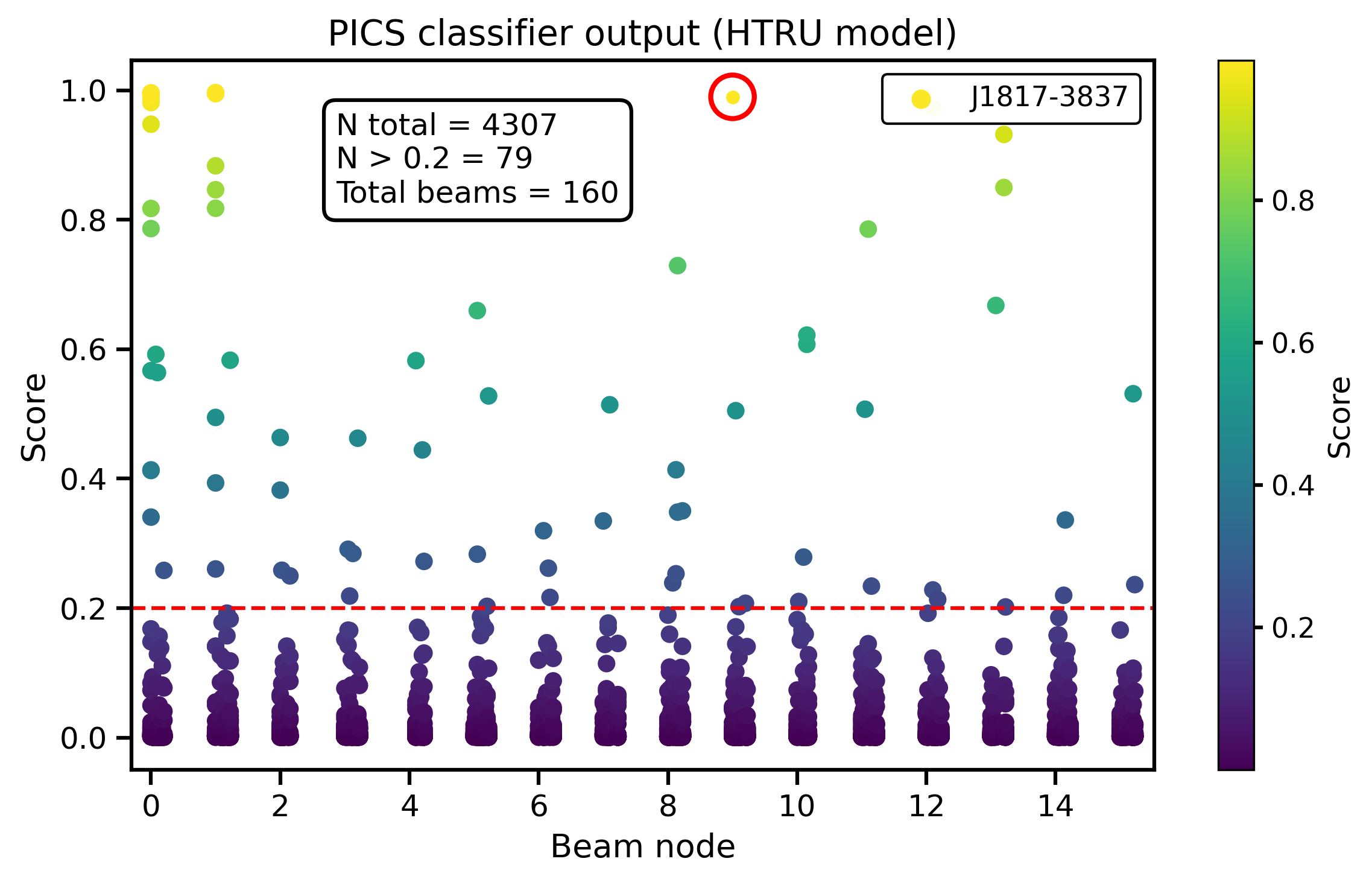}
}
\caption{Comparison of candidate distributions across different processing stages and machine-learning classification of the {\tt ver0} pipeline. The top-left panel shows the raw candidates from the Acceleration Search output (plotted with a 10,000X downsampling factor), while the top-right panel shows the filtered candidates after independent sifting of each beam's candidates (plotted with a 5X downsampling factor). The bottom-left panel shows the final candidates after spatial sifting of all beam candidates. In these three panels, candidates are displayed in the S/N–period plane. The bottom-right panel shows the {\tt PICS} classifier scores as a function of beam node, where each node produces 10 beams, resulting in a total of 160 beams from 16 nodes. The adopted candidate-selection threshold (a probability threshold value of  0.2) is indicated by the horizontal red dashed line. Lastly, the low SNR detection of J1817$-$3837 from a far-away beam is shown through each stage.}
\label{fig:ver0_output_plots}
\end{figure*}

\subsection{The FFT-based multi-GPU pipeline}
\label{subsec:pipeline}
With the periodicity range and band-dependent DM plan defined above (see Section~\ref{subsec:search_strategy}), the data are subsequently processed through the periodicity search pipeline to identify new pulsar candidates. The pipeline has been specifically designed to efficiently handle large volumes of multibeam data within a processing timescale, ensuring that all accumulated SPOTLIGHT commensal data can be processed and cleared on a weekly basis. This processing is carried out using the available computational resources allocated for offline analysis (typically 10 GPUs with 240 CPUs), while preserving dedicated resources required for correlator and beamformer operations, as well as real-time FRB searches.

Following the commensal FRB search and subsequent beam dumping, a script (called {\tt xtract2fil}\footnote{See the GitHub repository: \url{https://github.com/nsmspotlight/xtract2fil.git}}) automatically converts the raw data into filterbank format, preserving the native time and frequency resolution (1.3 ms sampling with 4096 channels) for FFT-based pulsar searches. During this conversion, the script also generates a reduced dataset of 13 ms with 1024 channels, optimised for the earlier discussed FFA pipeline targeting slow pulsar searches. This approach enables the FFT and FFA pipelines to operate independently on datasets tailored to their respective sensitivity regimes.

For each scan, the 160 beamformed PC data are first processed using the RFI-mitigation tool {\tt FILTOOL} to identify and excise radio frequency interference (RFI). {\tt FILTOOL} is part of the {\tt PulsarX} \footnote{{\tt PulsarX} GitHub repositiory: \url{https://github.com/ypmen/PulsarX.git}} \citep{Men_2023} software package developed under the TRAPUM survey. This initial RFI-cleaning step is essential for preserving the survey sensitivity by minimising contamination from terrestrial signals. In addition, it significantly reduces the number of spurious candidates generated during the search process, thereby lowering the computational burden of the subsequent pipeline stages. The cleaned data are then passed to the GPU-accelerated processing module {\tt AstroAccelerate (AA)} \footnote{{\tt AstroAccelerate} GitHub repository: \url{https://github.com/AstroAccelerateOrg/astro-accelerate.git}} \citep{Carels_2019} for further analysis. Developed under the Indo–UK collaboration, {\tt AstroAccelerate} is designed to process time-domain data efficiently on GPU architectures. In its current implementation, {\tt AA} performs periodicity searches (including dedispersion, FFT and 1-D harmonic summing up to the faintest detectable harmonics with a maximum of 32) independently for each beam over the predefined DM plan described in Section~\ref{subsec:search_strategy}. Future developments will incorporate Fourier domain acceleration searches with 2-D harmonic summing into {\tt AA}, which are currently in the testing phase (see Section~\ref{subsec:processing_and_future_dev}).

After the periodicity search, all candidates for each beam are independently sifted based on DM clustering, along with threshold cuts on S/N (signal-to-noise ratio) and harmonic number. This same sifting logic is also implemented in the single-beam pipeline, namely the PSS\footnote{{\tt PSS} GitHub repository: \url{https://github.com/jyotirmoydas5392/Pulsar_Search_Script.git}}, developed under the GCGPS\footnote{The GCGPS survey website: \url{https://www.ncra.tifr.res.in/~jroy/GC.html}} project \citep{Das_2025, Das_2026} for uGMRT data. The first two subplots (Figure~\ref{fig:ver0_output_plots}, a and b) illustrate the significant reduction in the number of candidates after applying the above filtering criteria. At this stage, more than 99.99\% of the candidates are rejected from further analysis. Despite this substantial reduction, the filtering process remains sensitive to low-S/N true pulsar candidates. As seen in subplot~b, candidates with S/N as low as $\sim3\sigma \:-5\sigma$ can still be retained if they exhibit high harmonic numbers, making them viable for subsequent processing.

\begin{figure*}
\centering

\subfigure[Spatial clustering and sifting of candidates]{
\includegraphics[width=0.48\textwidth]{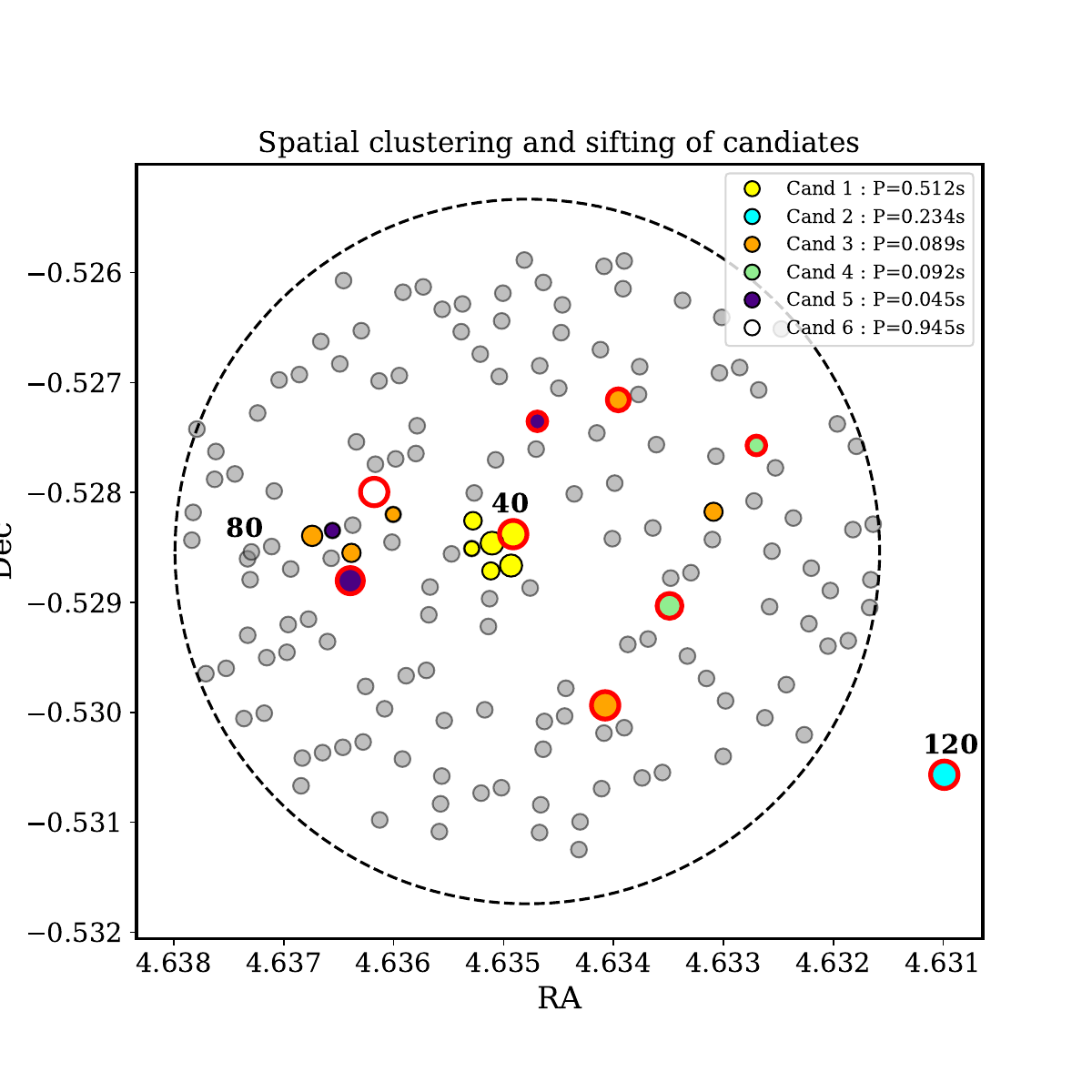}
}
\hfill
\subfigure[Harmonic + SNR ranking to sift isolated candidates in spatial clustering]{
\includegraphics[width=0.48\textwidth]{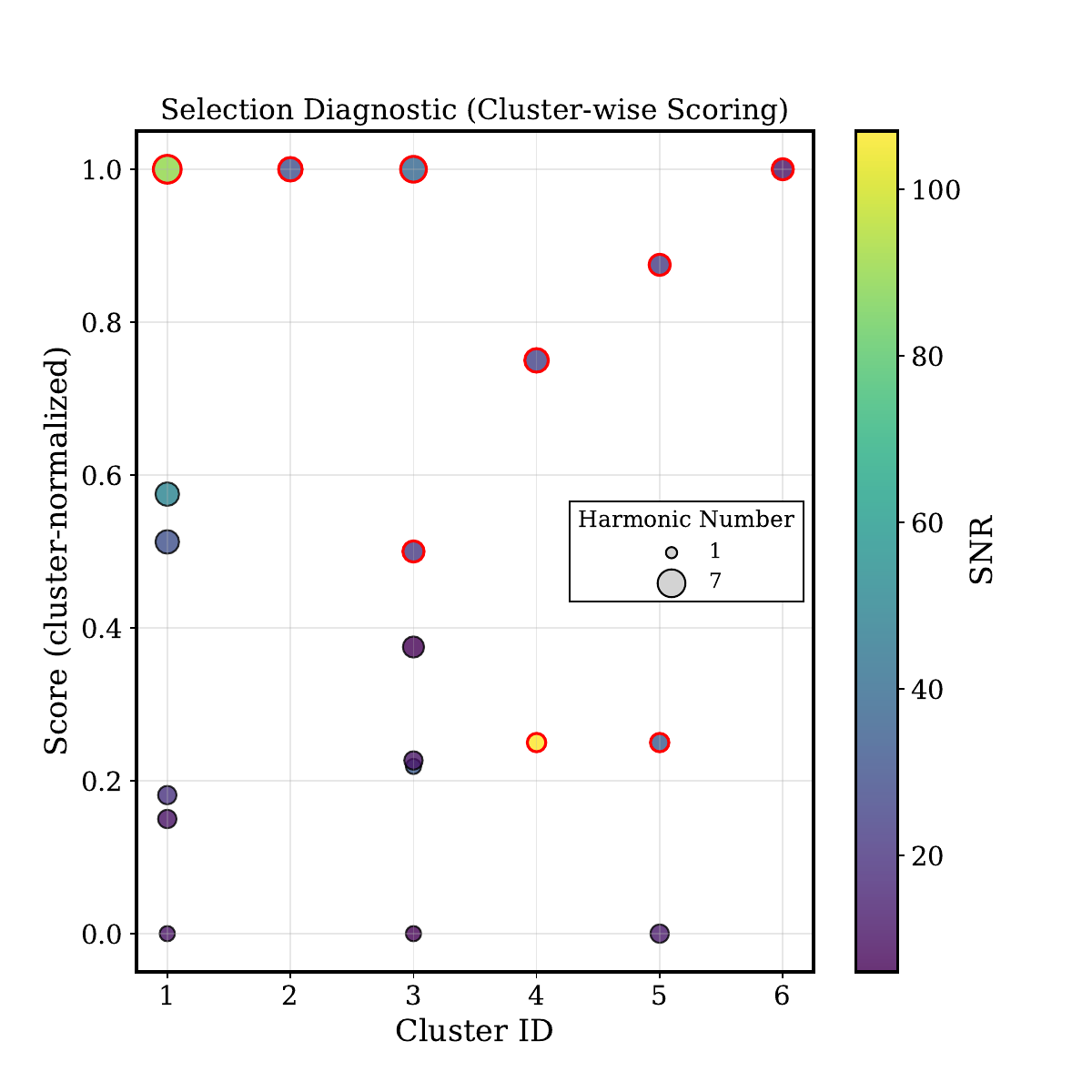}
}

\vspace{0.35cm}

\caption{Left: Spatial clustering of candidates in the RA--Dec plane for six representative periods. A prominent central cluster corresponds to a bright pulsar detection, while several isolated repetitions are distributed across the field. Cluster ID 6 originates from a single detection in Beam 120. Right: Candidate ranking for all identified clusters and repetitions. For clustered detections, only the highest-S/N candidate is retained, while isolated repetitions are ranked using a metric combining harmonic significance and S/N. Cluster ID 4 illustrates a case where a lower-S/N candidate with stronger harmonic content is ranked above a higher-S/N candidate with weaker harmonic significance.}
\label{fig:beam_clustering_visual}
\end{figure*}

\begin{figure*}
\centering
\includegraphics[width=\textwidth]{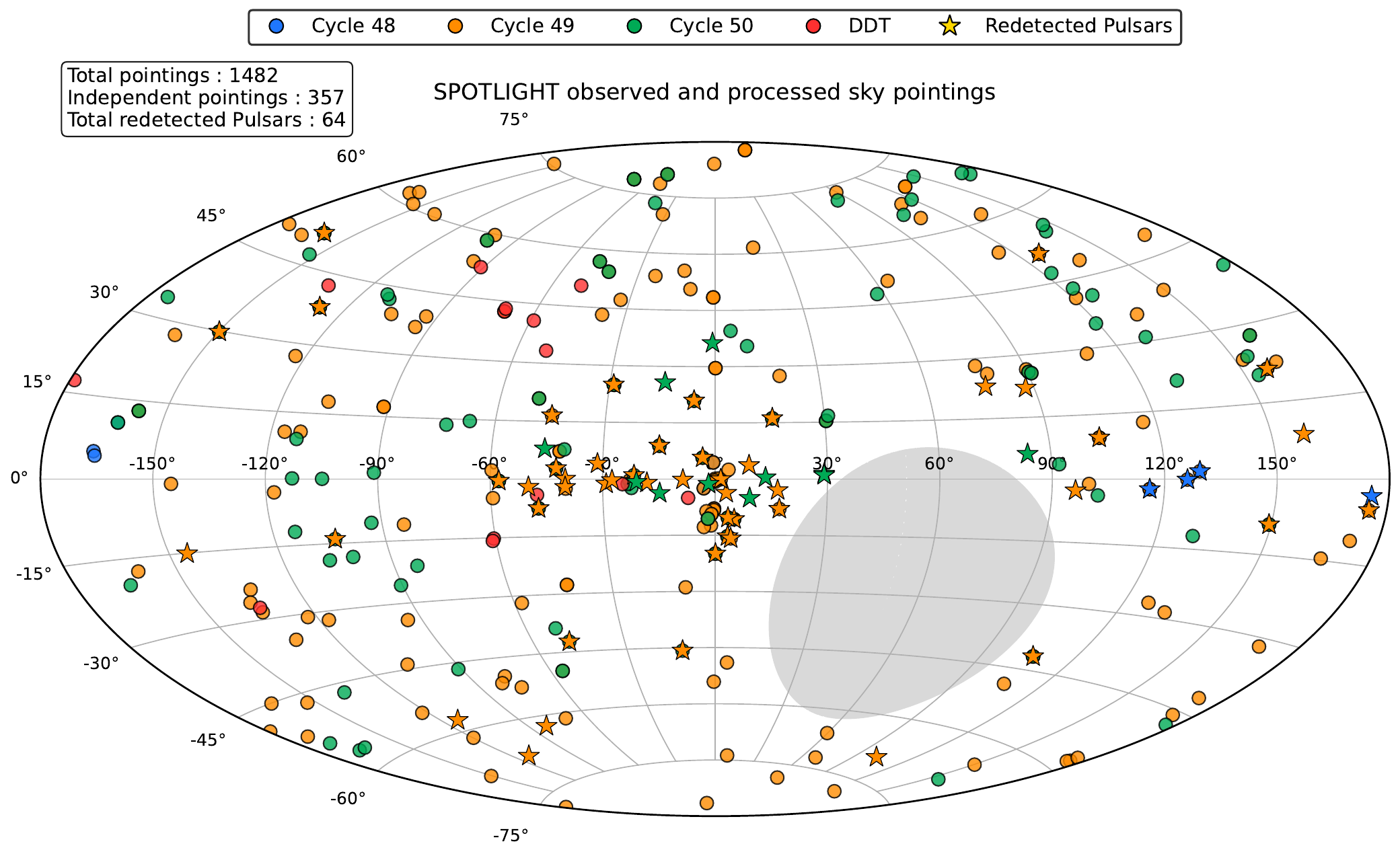}

\caption{Distribution of the 64 pulsars re-detected by the SPOTLIGHT {\tt ver0} FFT-pipeline from observations obtained during uGMRT Cycles 48--50 (as of May 2026). The locations of the re-detected pulsars are shown in Galactic coordinates together with the cumulative SPOTLIGHT commensal survey coverage. The shaded region indicates the sky area outside the uGMRT sky coverage ($\rm \leq -55\: degrees$), while star symbols mark the re-detected pulsars. Different colours represent observations from different observing cycles.}
\label{fig:redetected_pulsar_distribution}
\end{figure*}

Despite this efficient filtering, the number of surviving candidates remains too large to be processed directly through the computationally intensive folding stage. To further reduce the candidate set, we exploit the spatial information across all beams and correlate detections corresponding to the same candidate period. For each candidate period, we consider detections across all beams and group them based on spatial proximity, prioritising beams with higher S/N. Starting from the highest S/N beam, we iteratively identify detections in consecutive neighbouring beams to form clusters. For each such cluster, we retain only the detection from the beam with the highest S/N as the representative candidate. An example of such clustering is shown in Figure~\ref{fig:beam_clustering_visual}(a) and (b), where a group of detections around beam ID 40 (Cluster ID 1) is reduced to a single candidate corresponding to the highest S/N beam, which is then selected for subsequent folding. For detections that do not form consecutive beam clusters (i.e., isolated detections), we apply a ranking algorithm. These candidates are ranked using a weighted score, with 75\% weight assigned to the harmonic number and 25\% weight to the detection S/N. From this group, we select the top 5\% of candidates, with a minimum of two candidates retained. In addition, candidates detected in only one or two beams are retained regardless of their S/N or harmonic number, ensuring that potentially weak pulsars or sources located far from the beam centres are not inadvertently discarded. Examples of such cases are shown in Figure~\ref{fig:beam_clustering_visual}(b), including detections at beam ID 120 and Cluster ID 6. Applying this procedure across all candidate periods results in a typical reduction of candidates by a factor of 5–10 (see subfigures b and c of Figure \ref{fig:ver0_output_plots}); still, the final candidate set remains sensitive to both high and low S/N pulsars and is subsequently passed to the folding stage.

For the folding stage, we use the CPU-based folding package {\tt PSRFOLD}, which comes under the same software {\tt PulsarX}. This package processes all candidates simultaneously and employs a pruning-tree dedispersion algorithm, making it significantly faster than the classical {\tt PRESTO prepfold} approach.

Following the folding stage, candidate classification is performed using the AI-based classifier {\tt PICS}\footnote{{\tt PICS} GitHub repository: \url{https://github.com/zhuww/ubc_AI.git}} \citep{Zhu_2014}. This framework is compatible with folding outputs from both {\tt PRESTO prepfold} (PFD files) and {\tt PulsarX} (AR files), allowing seamless integration into our pipeline. In the SPOTLIGHT pipeline, we employ the pre-trained {\tt HTRU} model within {\tt PICS} and adopt a probability threshold of 0.2 for candidate selection (see Figure~\ref{fig:ver0_output_plots}d). This threshold is chosen to effectively suppress noise-dominated foldings while retaining a high completeness for real pulsar candidates. As illustrated in Figure~ \ref{fig:ver0_output_plots}d, this step results in a substantial reduction in the number of candidates (keeping true positives with manageable false positives), with a clear separation between noise and likely astrophysical signals. Consequently, the final set of candidates is reduced to $\sim$5–10\% of the folded candidates, making manual inspection both manageable and efficient. As a future development, after accumulating a substantial number of confirmed pulsar detections, we plan to train a dedicated GMRT-specific model to optimise classification performance in the SPOTLIGHT pipeline further.

All the steps described above are integrated into a unified pipeline, which we refer to as {\tt ver0}\footnote{{\tt ver0} GitHub repository: \url{https://github.com/jyotirmoydas5392/spotlight_pulsar_search_pipeline.git}}. Figure~\ref{fig:pipeline_block_diagram} illustrates the block diagram of the main processing stages of {\tt ver0}. This pipeline is currently deployed for routine pulsar searches on the 160 tied-array beam data generated by the SPOTLIGHT backend. Now, to process the simultaneously recorded IA beam alongside the 160 PC beams processed with {\tt ver0}, we follow the same core pipeline steps described above, with the exception of beam-level clustering, as spatial correlation across beams is not applicable to a single IA beam. This dedicated single-beam processing pipeline, named {\tt IApipe}, has already been deployed for continuous processing of the IA-beam data. By exploiting the substantially larger field of view of the IA beam, {\tt IApipe} enhances the survey's discovery potential (see Figure~\ref{fig:PsrPopPy2_outputs}) for relatively bright pulsars located at large angular offsets from the phase centre that may lie outside the PC beam coverage.

\subsection{Validation and preliminary results}
\label{subsec:validation_results}
The SPOTLIGHT pulsar search pipeline {\tt ver0} has been operational since mid-2025 (uGMRT Cycle 48). Since its deployment, the pipeline has undergone extensive testing and optimisation to ensure robust processing of large data volumes while maintaining sensitivity to low-S/N candidates throughout the filtering and candidate-sifting stages (as discussed in Section \ref{subsec:pipeline}).

We validated the pipeline using a large number of commensal observations containing known pulsars. Through extensive testing and iterative development, significant improvements were made to the candidate filtering, folding, and classification stages. As a result, the current version of the pipeline is capable of reliably detecting pulsars and promising candidates down to detection significances of approximately $4$--$5,\sigma$. To date, across all three commensal uGMRT bands, we have covered $\sim$ 245.92 sq. degrees of uGMRT sky-searching for periodic radio signals between July 2025 and May 2026, successfully re-detecting several known pulsars from SPOTLIGHT commensal observations. Figure~\ref{fig:redetected_pulsar_distribution} presents the pulsars re-detected by the {\tt ver0} pipeline across multiple uGMRT observing cycles, illustrating the steady improvement in survey performance over time.

\begin{figure}
\centering
    \includegraphics[width=\columnwidth]{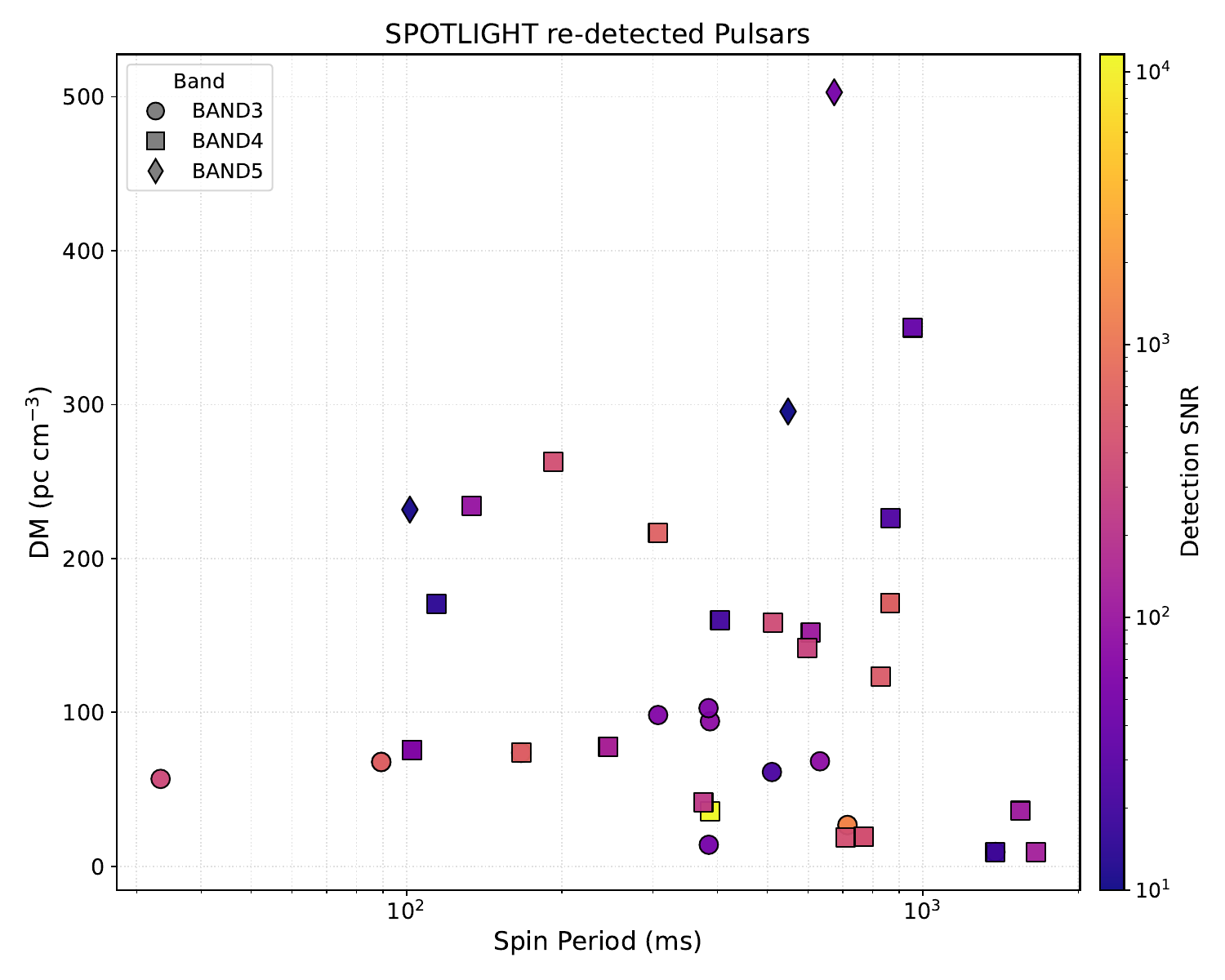}
    \caption{This figure presents a subset of the pulsars (spin period proportional to marker size) re-detected in SPOTLIGHT observations using the {\tt ver0} processing pipeline, primarily from uGMRT Cycle 49. This presents detections over a diverse range of Period and Dispersion Measure (DM) spans from all three SPOTLIGHT commensal uGMRT bands, from the FFT processing.}
    \label{fig:SPOTLIGHT_redetected_pulsars}
\end{figure}

During uGMRT Cycle 48, the overall performance was limited by non-optimal beam scaling in the correlator/beamformer system, while the pulsar search pipeline was still undergoing development, particularly in its RFI mitigation and candidate filtering stages. These factors reduced the sensitivity to faint pulsars near the detection threshold, resulting in very few successful detections, as also evident from Figure~\ref{fig:redetected_pulsar_distribution}. By Cycle 49, improvements in the beamforming system, together with substantial advances in RFI excision, candidate sifting, and classification algorithms in {\tt ver0}, led to a significant increase in sensitivity and a corresponding rise in the number of reliable pulsar detections. Figure~\ref{fig:SPOTLIGHT_redetected_pulsars} presents a representative subset of these detections, primarily from the latter part of Cycle~49. The successful detection of pulsars spanning a wide range of spin periods, dispersion measures, and flux densities demonstrates the robustness of the pipeline and its readiness for the efficient discovery of new pulsars.

The absence of new pulsar discoveries so far is therefore consistent with the reduced sensitivity during the early phases of the survey, rather than any limitation of the survey strategy itself. With the beamforming and pipeline optimisations now in place, the survey sensitivity has improved substantially, and several promising low-S/N ($\sim$ 5$\sigma$ to 12-13$\sigma$) pulsar candidates have already been identified from late Cycle 49 and currently ongoing Cycle 50 observations. Dedicated follow-up observations for these candidates are currently underway. We expect to report on the confirmation and detailed characterisation of these candidates in future publications arising from the SPOTLIGHT survey.

\subsection{Processing speed and future developments}
\label{subsec:processing_and_future_dev}
Following the required optimisations, the current {\tt ver0} pipeline is now operating faster than real-time performance for typical processing on 10 GPUs in 5 GPU nodes. Across all observing bands, DM search configurations, and integration times, the average processing time is compatible with routine survey operations. In practice, the accumulated data can be processed and cleared weekly without creating disk-space bottlenecks, enabling seamless long-term survey execution. To quantify the upper limit of the computational performance, a typical 40-minute observation (representative of a standard continuum scan) requires approximately an hour to process in a total of 5 GPU nodes, when searched over the full parameter space up to the instrumental DM limits described in Section~\ref{subsec:search_strategy}. The processing time is largely independent of the observing band: Band 3 observations require searches to relatively low maximum DMs but with finer DM spacing, whereas Band 5 observations extend to much higher DMs with correspondingly coarser DM steps, resulting in a comparable overall computational load. A typical Band 3 observation produces hundreds of million candidates after the periodicity search stage, with somewhat fewer candidates in Bands 4 and 5. Following the application of the candidate filtering and classification framework, this number is reduced to a few hundred candidates per observation for manual verification, depending on the presence of strong known pulsars, the prevailing RFI environment, and the overall quality of the dataset.

About the ongoing and future developments, the first major upgrade to the pipeline will be the inclusion of a Fourier domain acceleration search with 2-D harmonic summing.  An {\tt astro-accelerate} module for acceleration searches up to a predefined number of Fourier bin drifts, together with 2D harmonic summing, is already in place. The code is currently undergoing final testing and validation before integration into the {\tt ver0} pipeline. This enhancement will make the pipeline sensitive to mildly recycled binary millisecond pulsars (MSPs), as well as double neutron star (DNS) systems where measurable line-of-sight acceleration is present during the observation.

Now, coming to the beam selection part, at present, 160 PC beams are randomly selected and recorded out of the 640 operational beams available for real-time FRB searches. This random beam-selection scheme introduces a non-negligible reduction in survey completeness, as pulsars residing within non-recorded beams have a substantial probability of escaping detection despite being present within the surveyed field of view. Following the integration of acceleration search capabilities, we plan to optimise the beam selection and data-dumping logic to prioritise beams more likely to contain periodic signals.
This optimisation will be facilitated by the newly developed module {\tt PULSCAN}\footnote{GitHub repository: \url{https://github.com/jack-white1/pulscan}} \citep{White_2025}. {\tt PULSCAN} is a highly optimised GPU-based code that employs a fast boxcar-based algorithm to scan FFT power spectra for periodic signals. Owing to its efficient implementation, the code operates significantly faster than real-time for a reasonable DM plan. We plan to deploy {\tt PULSCAN} during real-time FRB observations, where all 640 beam data streams reside in shared memory. {\tt PULSCAN} will iterate through these beams, rank them according to the probability of containing a periodic signal, and dynamically select the 160 most promising beams for data dumping.

We have already rigorously tested {\tt PULSCAN} in a pseudo-shared-memory framework using all 640 beams and the intended DM plans, confirming that it can process SPOTLIGHT shared-memory data faster than real time while effectively ranking beams. This intelligent beam selection strategy will substantially reduce the probability of discarding beams containing potentially detectable pulsars without inspection, thereby maximising discovery efficiency. We aim to implement this system immediately following the integration of the acceleration-search upgrade within the {\tt AstroAccelerate} framework.

\section{Summary}
\label{sec:summary}
In this paper, we present the FFT-based pulsar search pipeline developed for the SPOTLIGHT commensal survey. SPOTLIGHT is a commensal uGMRT backend designed primarily for real-time detection and localisation of fast radio bursts (FRBs) from ongoing uGMRT observations in a commensal way. In addition to its primary objective of real-time FRB detection, SPOTLIGHT records a selected subset of the beamformed data products (currently 160 PC beams), which are subsequently processed offline for periodicity transient searches. To efficiently process this large volume of data for periodic searches, we have developed an {\tt AstroAccelerate}- based multi-GPU FFTsearch pipeline called {\tt ver0} capable of handling high data rates and ensuring that all recorded data are processed on a weekly timescale, thereby maintaining a steady flow between data acquisition and analysis throughout the commensal observing cycles.

We first provide an overview of the SPOTLIGHT backend, from a brief description of the beamformer followed by the correlator and data recording. We discuss how SPOTLIGHT is optimally covering its desired Field of View (FoV) and known transients present in its FoV with the currently formed 640 PC beams and the simultaneously recorded IA beam, along with subsequent dumping of a total of 160 PC beams (containing all the infilled source-targeted PC beams) and the IA beam as well.

We discuss our motivation behind carrying out this pulsar search survey, along with a detailed work on the discovery estimation using the {\tt PsrPopPy2} model and three years (total six uGMRT Cycles) of uGMRT sky coverage details, where we estimate that with a total of 160 recorded PC beams and one IA beam, the SPOTLIGHT system is capable of discovering $\sim 450$ new pulsars over seven uGMRT cycles having three and half years of commensal observations. Subsequently, we describe the search parameter space, limited by the data sampling rate and FFT sensitivity loss for long-period pulsars. Then we describe the {\tt AstroAccelerate}- based multi-GPU FFT search pipeline in detail, including its performance, candidate optimisation strategies, subsequent folding and candidate classification performance. We also present the works validating the pipeline, the redetections of known pulsars over Cycles 48 to 50, the sensitivity improvement over time, and the current state of the pipeline. Finally, we discuss ongoing and future developments aimed at improving the sensitivity, efficiency, and overall scientific impact of the pulsar search component within the SPOTLIGHT survey.

\begin{acknowledgments}
We acknowledge the invaluable contribution of colleagues from the Oxford e-Research Centre (OeRC), NVIDIA, and the Centre for the Development of Advanced Computing (C-DAC). 
We thank our colleagues at the Centre for Development of Advanced Computing (C-DAC) for their support in setting up the Param Brahmand data centre at the GMRT. 
We acknowledge the funding of the SPOTLIGHT backend (called Param Brahmand) under the National Supercomputing Mission's (NSM) Phase 3, as well as the support of the Department of Atomic Energy (DAE), Government of India, under project no. $\rm 12-R\&D-TFR5.02-0700$, for the contributions towards the overhead cost. The uGMRT is operated by the National Centre for Radio Astrophysics of the Tata Institute of Fundamental Research, India. We gratefully acknowledge support from the \enquote{Building Indo–UK Collaborations Towards the Square Kilometre Array}, funded under the DAE–STFC Technology and Skills Programme, which facilitated the development of the highly efficient GPU-based pulsar search pipeline. We sincerely thank the uGMRT engineers involved in the SPOTLIGHT survey for their relentless efforts in commissioning and stabilising the correlator and beamformer systems, ensuring reliable and high-quality data acquisition in the commensal mode for offline pulsar search under the SPOTLIGHT survey. We also thank the uGMRT operators for their coordinated efforts in successfully conducting the SPOTLIGHT survey observations. This research was supported in part by the International Centre for Theoretical Sciences (ICTS) for the FTSky: A program in the field of Fast Radio Transients (code: ICTS/FTSky2025/10).
\end{acknowledgments}


\bibliography{bibliography}{}
\bibliographystyle{aasjournal}

\end{document}